\documentclass[twocolumn,showpacs,amsmath,amssymb,floatfix]{revtex4}

\usepackage{graphicx}% Include figure files
\usepackage{dcolumn}% Align table columns on decimal point
\usepackage{bm}% bold math
\begin{document}

\title{The $^{18}$Ne($\alpha$,$p$)$^{21}$Na breakout reaction in x-ray bursts: experimental determination of
spin-parities for $\alpha$ resonances in $^{22}$Mg via resonant elastic scattering of $^{21}$Na+$p$}

\author{J.J.~He$^1$}
\email{jianjunhe@impcas.ac.cn}
\author{L.Y.~Zhang$^{1,2,3}$}
\author{A.~Parikh$^{4,5}$}
\author{S.W.~Xu$^{1}$}
\author{H.~Yamaguchi$^6$}
\author{D.~Kahl$^{6}$}
\author{S.~Kubono$^{1,6,11}$}
\author{J.~Hu$^{1}$}
\author{P.~Ma$^{1}$}
\author{S.Z.~Chen$^{1,3}$}
\author{Y.~Wakabayashi$^{7,11}$}
\author{B.H.~Sun$^{8}$}
\author{H.W.~Wang$^{9}$}
\author{W.D.~Tian$^{9}$}
\author{R.F.~Chen$^{1}$}
\author{B.~Guo$^{10}$}
\author{T.~Hashimoto$^{6}$}
\author{Y.~Togano$^{11}$}
\author{S.~Hayakawa$^{6}$}
\author{T.~Teranishi$^{12}$}
\author{N.~Iwasa$^{13}$}
\author{T.~Yamada$^{13}$}
\author{T.~Komatsubara$^{14}$}

\affiliation{$^1$Institute of Modern Physics, Chinese Academy of Sciences, Lanzhou 730000, China}
\affiliation{$^2$School of Nuclear Science and Technology, Lanzhou University, Lanzhou 730000, China}
\affiliation{$^3$University of Chinese Academy of Sciences, Beijing 100049, China}
\affiliation{$^4$Departament de F\'{\i}sica i Enginyeria Nuclear, EUETIB, Universitat Polit\`{e}cnica de Catalunya, Barcelona E-08036, Spain}
\affiliation{$^5$Institut d'Estudis Espacials de Catalunya, Barcelona E-08034, Spain}
\affiliation{$^6$Center for Nuclear Study, the University of Tokyo, RIKEN campus, Wako, Saitama 351-0198, Japan}
\affiliation{$^7$Advanced Science Research Center, Japan Atomic Energy Agency (JAEA), Ibaraki 319-1106, Japan}
\affiliation{$^8$School of Physics and Nuclear Energy Engineering, Beihang University, Beijing 100191, China}
\affiliation{$^9$Shanghai Institute of Applied Physics, Chinese Academy of Sciences, Shanghai 201800, China}
\affiliation{$^{10}$China Institute of Atomic Energy, P.O. Box 275(46), Beijing 102413, China}
\affiliation{$^{11}$RIKEN (The Institute of Physical and Chemical Research), Wako, Saitama 351-0198, Japan}
\affiliation{$^{12}$Department of Physics, Kyushu University, 6-10-1 Hakozaki, Fukuoka 812-8581, Japan}
\affiliation{$^{13}$Department of Physics, University of Tohoku, Miyagi 980-8578, Japan}
\affiliation{$^{14}$Department of Physics, University of Tsukuba, Ibaraki 305-8571, Japan}

\date{\today}

\begin{abstract}
The $^{18}$Ne($\alpha$,$p$)$^{21}$Na reaction provides a pathway for breakout from the hot CNO cycles to the 
$rp$-process in type I x-ray bursts. To better determine this astrophysical reaction rate, the resonance parameters of the compound
nucleus $^{22}$Mg have been investigated by measuring the resonant elastic scattering of $^{21}$Na+$p$. An 89 MeV $^{21}$Na
radioactive ion beam was produced at the CNS Radioactive Ion Beam Separator and bombarded an 8.8 mg/cm$^2$ thick polyethylene target.
The recoiled protons were measured at scattering angles of $\theta_{c.m.}$$\approx$175${^\circ}$ and 152${^\circ}$ by three
$\Delta E$-$E$ silicon telescopes. The excitation function was obtained with a thick-target method over energies
$E_x$($^{22}$Mg)=5.5--9.2 MeV. The resonance parameters have been determined through an $R$-matrix analysis. For the first time,
the $J^{\pi}$ values for ten states above the alpha threshold in $^{22}$Mg have been experimentally determined in a single consistent
measurement. We have made three new $J^{\pi}$ assignments and confirmed seven of the ten tentative assignments in the previous work.
The $^{18}$Ne($\alpha$,$p$)$^{21}$Na reaction rate has been recalculated, and the astrophysical impact of our new rate
has been investigated through one-zone postprocessing x-ray burst calculations. We find that the $^{18}$Ne($\alpha$,$p$)$^{21}$Na
rate significantly affects the peak nuclear energy generation rate and the onset temperature of this breakout reaction in these
phenomena.
\end{abstract}


%\keywords{Reaction induced by unstable nuclei, Decay by proton emission, nuclear astro}

\maketitle

% main text
Type I x-ray bursts (XRBs), one of the most fascinating astrophysical phenomena, are characterized by sudden dramatic increases in
luminosity of roughly 10--100 s in duration, with a total energy release of about 10$^{39}$ erg per burst. These recurrent phenomena
(on timescales of hours to days) have been the subject of many observational, theoretical and experimental studies (for reviews see
{\it e.g.},~\cite{bib:lew93,bib:str06,bib:par13}). The characteristics of XRBs have been surveyed extensively in a number of
space-borne x-ray satellite observatory missions, including RXTE, BeppoSAX, Chandra, HETE-2, and XMM/Newton. More than 90 galactic
XRBs have been identified since their initial discovery in 1976. These observations have provided abundant data and opened a new era
in x-ray astronomy. The bursts have been interpreted as being generated by thermonuclear runaway on the surface of a neutron star
that accretes H- and He-rich material from a less evolved companion star in a close binary system~\cite{bib:woo76,bib:jos77}. The
accreted material burns stably through the hot, $\beta$-limited carbon-nitrogen-oxygen (HCNO)~\cite{bib:laz79,bib:wie99} cycles,
giving rise to the persistent flux. Once critical temperatures and densities are achieved, breakout from this region can occur
through, {\it e.g.}, $\alpha$-induced reactions on the waiting point nuclei $^{14}$O, $^{15}$O and $^{18}$Ne. Through the rapid
proton capture process (rp-process)~\cite{bib:wal81,bib:sch98,bib:woo04}, this eventually results in a rapid increase in energy
generation (ultimately leading to the XRB) and nucleosynthesis up to A$\sim$100 mass region~\cite{bib:sch01,bib:elo09}. Among the
possible breakout reactions, breakout may occur through the $^{18}$Ne($\alpha$,$p$)$^{21}$Na reaction~\cite{bib:wie99}; however, the 
actual astrophysical conditions under which this occurs depend critically on the actual $^{18}$Ne($\alpha$,$p$)$^{21}$Na 
thermonuclear rate. Over stellar temperatures achieved in XRBs, this rate has not been sufficiently well determined.

The reaction rate for $^{18}$Ne($\alpha$,$p$)$^{21}$Na is dominated by contributions from resonances in the compound nucleus
$^{22}$Mg above the $\alpha$ threshold at 8.142 MeV~\cite{bib:wan12}. The temperature region of interest in XRBs is about 0.4--2.0 GK,
corresponding to an excitation region of $E_x$=8.6--11.0 MeV in $^{22}$Mg. The first theoretical estimate~\cite{bib:gor95} of this
reaction rate was made based on rather limited experimental level-structure information in $^{22}$Mg. The uncertainty of this rate
was largely due to uncertainties in both excitation energies $E_x$ (or resonance energies $E_R$) and resonance strengths
$\omega \gamma$. After that, the levels in $^{22}$Mg have been extensively studied, and more than 40 levels were observed above the
$\alpha$ threshold. Such high level-density suggests that a statistical-model approach might provide a reliable estimate of the rate.
However, only natural-parity states in $^{22}$Mg can be populated by the $^{18}$Ne+$\alpha$ channel, and thus the effective level
density will be considerably lower. The $\alpha$-unbound states in $^{22}$Mg were previously studied by many transfer reaction
experiments. In the $^{12}$C($^{16}$O,$^6$He)$^{22}$Mg~\cite{bib:che01}, $^{25}$Mg($^3$He,$^6$He)$^{22}$Mg~\cite{bib:cag02} and
$^{24}$Mg($^4$He,$^6$He)$^{22}$Mg~\cite{bib:ber03} experiments, the excitation energies in $^{22}$Mg were determined with a typical
uncertainty of $\pm$20--30 keV. Later on, the excitation energies were determined precisely by a
$^{24}$Mg($p$,$t$)$^{22}$Mg~\cite{bib:mat09} experiment, where the uncertainty of about 1--15 keV was achieved for most states above
the $\alpha$ threshold. With these precise energies, the uncertainties in $^{18}$Ne($\alpha$,$p$)$^{21}$Na rate can be largely reduced.
A new reaction rate was recommended based on the combined analysis of all available data~\cite{bib:mat13}, and just published
during the refereeing process for this article.

The above indirect studies mainly focused on the determination of excitation energies, and the spin-parity assignments were not
strictly constrained. Some spin-parity assignments were made~\cite{bib:gor95,bib:che01,bib:mat09} simply by referring to those of
mirror states in $^{22}$Ne; such assignments are dubious due to the high level-density in this excitation energy region. In a later
$^{24}$Mg($p$,$t$)$^{22}$Mg~\cite{bib:cha09} experiment, several spin-parity assignments were made via an angular distribution
measurement. However, the insufficient resolution of their measurements at the center-of-mass ($c.m.$) scattering angles of
$\theta_{c.m.}$ above 20$^{\circ}$ makes such $J^{\pi}$ assignments questionable~\cite{bib:mat09}. In addition, two tentative
spin-parity assignments were made in a previous low statistics measurement~\cite{bib:he08,bib:he09} of resonant $^{21}$Na+$p$
elastic scattering, and such assignments still need to be confirmed by a high statistics experiment.

A comparison of all available reaction rates shows discrepancies of up to several orders of magnitude around
$T$$\sim$1 GK~\cite{bib:mat09}, and therefore it remains unclear whether the statistical-model calculations provide a reliable rate
estimation in a wide temperature region. There are still many resonances (above the $\alpha$ threshold) without firm spin-parity
assignments, which need to be determined experimentally. As a consequence, the accuracy of the current
$^{18}$Ne($\alpha$,$p$)$^{21}$Na reaction rate is mainly limited by the lack of experimental spin-parity and spectroscopic
information of the resonances in $^{22}$Mg above the $\alpha$ threshold.

So far, only two direct measurements~\cite{bib:bra99,bib:gro02} for the $^{18}$Ne($\alpha$,$p$)$^{21}$Na reaction have been performed.
The lowest energies achieved in these studies ($E_{c.m.}$=2.0 and 1.7 MeV) are still too high compared with the energy region
$E_{c.m.}$$\leq $1.5 MeV of interest for HCNO breakout in XRBs. New results~\cite{bib:sal12} have recently become available at
the ISAC II facility at TRIUMF, where the $^{18}$Ne($\alpha$,$p$)$^{21}$Na cross section was determined in the energy region of
$E_{c.m.}$=1.19--2.57 MeV by measuring the time-reversal reaction $^{21}$Na($p$,$\alpha$)$^{18}$Ne in inverse kinematics.
Nonetheless, these results are still insufficient for a reliable rate at all temperatures encountered within XRBs.

In this work, the $^{18}$Ne($\alpha$,$p$)$^{21}$Na rate is determined via a new measurement of the resonant elastic scattering of
$^{21}$Na+$p$. This is an entirely new high-statistics experiment comparing to the previous one~\cite{bib:he08,bib:he09}. In the
resonant elastic-scattering mechanism, $^{22}$Mg is formed via sub-Coulomb barrier fusion of $^{21}$Na+$p$ as an excited compound
nucleus, whose states promptly decay back into $^{21}$Na+$p$. This process interferes with Coulomb scattering resulting in a
characteristic resonance pattern in the excitation function~\cite{bib:rui05}. With this approach, the excitation function was
obtained simultaneously in a wide range of 5.5--9.2 MeV in $^{22}$Mg with a well-established thick-target
method~\cite{bib:art90,bib:gal91,bib:kub01}. For the first time, we have experimentally determined the $J^{\pi}$ values for ten
states above the $\alpha$ threshold in $^{22}$Mg.

The experiment was performed using CRIB (CNS Radioactive Ion Beam separator) at the Center for Nuclear Study (CNS) of the University
of Tokyo~\cite{bib:kub02,bib:yan05}. During the last decade, the radioactive ion beams (RIBs) produced at CRIB have been successfully
utilized in the resonant scattering experiments with a thick-target method~\cite{bib:ter03,bib:ter07,bib:hjj08,bib:yam09,bib:he09},
which proved to be a successful technique as adopted in the present study. Some details about this experiment were preliminarily
described elsewhere~\cite{bib:omega,bib:nic}. An 8.2 MeV/nucleon primary beam of $^{20}$Ne$^{8+}$ was accelerated by an AVF cyclotron
($K$=79) at RIKEN, and bombarded a liquid nitrogen-cooled $D_{2}$ gas target (90 K)~\cite{bib:yam08} with an average intensity of
65 pnA. The thickness of $D_{2}$ gas was about 2.9 mg/cm$^{2}$ at 530 Torr pressure. The $^{21}$Na beam was produced via the
$^{20}$Ne($d$,$n$)$^{21}$Na reaction in inverse kinematics. After the Wien filter, a purity of 70\% for the $^{21}$Na beam was
achieved on the target.

Two parallel-plate avalanche counters (PPACs)~\cite{bib:kum01} measured the timing and position of the incoming beam. The beam
impinged on an 8.8 mg/cm$^2$ polyethylene (CH$_2$)$_n$ target, which was thick enough to stop all the beam ions. In addition, a
10 mg/cm$^2$ thick carbon target was used for evaluating the C background contribution. The $^{21}$Na beam bombarded the targets at
energy about 89.4 MeV ($\Delta E$=1.9 MeV in FWHM). The averaged beam intensity was about 2$\times$10$^{5}$ pps. The beam particles
were clearly identified in an event-by-event mode using position and timing signals~\cite{bib:omega,bib:nic}. The recoiled light
particles were detected with three Micron~\cite{bib:micron} silicon $\Delta E$-$E$ telescopes centered at angles of
$\theta_\mathrm{Si}$=0$^\circ$, +14$^\circ$ and -14$^\circ$ with respect to the beam line, respectively. Each $\Delta E$-$E$
telescope subtended an opening angle of about 10$^\circ$ with a solid angle of about 27 msr in the laboratory frame. In the
$c.m.$ frame for elastic scattering, the relevant averaged scattering angles are determined to be
$\theta_{c.m.}$$\approx$175$^\circ$, 152$^\circ$ and 151$^\circ$, respectively. The double-sided-strip (16$\times$16 strips)
$\Delta E$ detectors measured the energy, position and timing signals of the particles, and the pad $E$ detectors measured their
residual energies. This allowed for the clear identification of recoiled particles~\cite{bib:omega,bib:nic}. The energy calibration
for the Si detectors was carried out by using secondary proton beams produced with CRIB and a standard triple-$\alpha$ source.

\begin{figure}
\includegraphics[width=8.6cm]{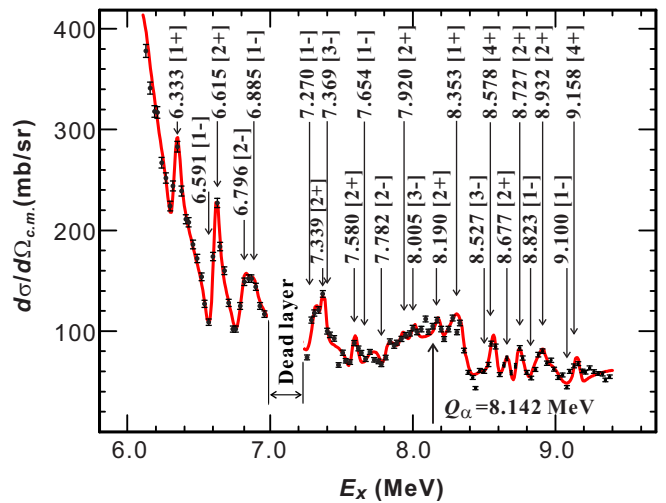}
\vspace{-6mm}
\caption{\label{fig1} (Color online) Experimental $c.m.$ differential cross section for resonant elastic scattering of $^{21}$Na+$p$
at a scattering angle of $\theta_{c.m.}$$\approx$175$^\circ$. It also shows a best overall $R$-matrix fit.}
\end{figure}

The $^{21}$Na+$p$ elastic-scattering excitation functions were reconstructed using the procedure described 
previously~\cite{bib:hjj08,bib:he09}.
Figure~\ref{fig1} shows the proton elastic-scattering spectrum for a scattering angle of $\theta_{c.m.}$$\approx$175${^\circ}$.
The cross-section data were corrected for the stopping cross sections of ions in the target~\cite{bib:kub01,bib:zie85}, and the
data within the dead-layer region (between $\Delta E$ and $E$ detectors) were removed from the figure. 
The normalized proton yield with the C target, whose spectrum was flat and smooth as a function of energy, was less than 
about 1/5 of that with the (CH$_2$)$_n$ target. In Fig.~\ref{fig1}, the carbon-induced background was subtracted accordingly and the 
uncertainties shown are mainly of statistical origin. The excitation energies indicated on Fig.~\ref{fig1} are calculated by 
$E_{x}$=$E_{R}$+$Q_{p}$. Here, the resonance energy $E_{R}$ is determined by an $R$-matrix analysis (see below), and a proton 
separation energy of $Q_{p}$=5.504 MeV is adopted~\cite{bib:muk04,bib:wan12}. With this thick-target technique, the $E_{c.m.}$ energy 
uncertainty is approximately $\pm$(30--50) keV for those states above the $\alpha$ threshold based on a Monte-Carlo 
simulation~\cite{bib:he08,bib:he09}. The present excitation energies agree with those adopted in Ref.~\cite{bib:mat09} within the 
uncertainties (see Table I).

The $^{21}$Na+$p$ excitation function has been analyzed by a multichannel $R$-matrix \cite{bib:lan58} code {\tt MULTI}~\cite{bib:nel85}.
An overall $R$-matrix fit is also shown in Fig.~\ref{fig1}. A channel radius of $R_n$=1.35(1+21$^\frac{1}{3}$)
fm~\cite{bib:gor95,bib:che01} was adopted in the calculation. The successful reproduction of the well-known
states~\cite{bib:rui05,bib:mat09} at 6.333, 6.591, 6.615, and 6.796 MeV by the code (see Fig.~\ref{fig1}) provides confidence in the
present method. In this paper, we focus on determining the resonance parameters of those states above the $\alpha$ threshold in
$^{22}$Mg, which eventually determine the $^{18}$Ne($\alpha$,$p$)$^{21}$Na reaction rate. The resonance parameters for all observed
states will be published elsewhere in more detail~\cite{bib:zhang}.

\begin{figure}
\includegraphics[width=8.6cm]{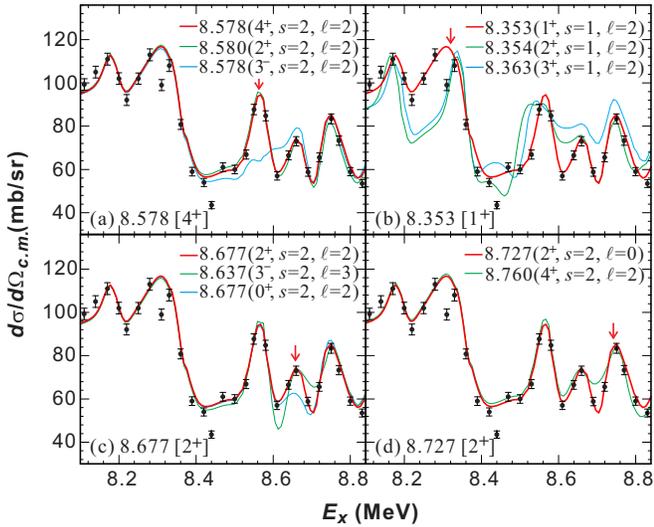}
\vspace{-6mm}
\caption{\label{fig2} (Color online) Sample $R$-matrix fitting results for some resonances above $\alpha$-threshold. The (red) thicker
lines represent the best fits. The relevant channel spin $s$ and orbital angular momentum $\ell$ values are indicated.}
\end{figure}

In total, ten resonances above the $\alpha$-threshold were observed and analyzed by the $R$-matrix code. For the first time, we have
experimentally confirmed the $J^{\pi}$ values tentatively assigned by Matic {\it et al.}~\cite{bib:mat09} for seven states at 8.181,
8.519, 8.574, 8.783, 8.932, 9.080 and 9.157 MeV, and assigned here new $J^{\pi}$ values for three states at 8.385, 8.657 and
8.743 MeV. As an example, the typical $R$-matrix fits with possible $J^\pi$, channel spin $s$ and orbital angular momentum $\ell$ for
the presently observed 8.578, 8.353, 8.677 and 8.727 MeV states are shown in Fig.~\ref{fig2}. The presently observed 8.578 MeV state
is closest to the 8.574 MeV in the work of Matic {\it et al.} in which it was assigned to be a 4$^+$ state based on a shell-model
calculation. As shown in Fig.~\ref{fig2}(a), both 2$^+$ and 4$^+$ can fit our data very well. As such, our data support the previous
4$^+$ assignment. The observed 8.353 MeV state is regarded as the 8.385 MeV state of Matic {\it et al.} whose $J^{\pi}$ was suggested
to be 2$^{+}$ by referring to the mirror state in $^{22}$Ne. In addition, we assigned it $J^{\pi}$=(1$^{+}$--3$^{+}$) in a previous
low statistics experiment~\cite{bib:he08,bib:he09} where 1$^+$ was also the most probable assignment. In this work, $J^{\pi}$=1$^{+}$
is again the best candidate as shown in Fig.~\ref{fig2}(b). Furthermore, this state was only weakly populated in the previous
transfer-reaction experiments~\cite{bib:che01,bib:ber03,bib:mat09} which preferentially populated the natural-parity states in
$^{22}$Mg. This again supports our assignment of 1$^+$ unnatural-parity to this state. The observed 8.677 MeV state corresponds to
the 8.657 MeV state of Matic {\it et al.}, which was assigned as a $J^{\pi}$=0$^{+}$ also based on a shell-model calculation.
However, such a prediction is questionable because of the high level density at such a high excitation energy region.
Matic {\it et al.} regarded this state as the 8.613 MeV state observed in Ref.~\cite{bib:che01} where it was assumed to be 3$^{-}$
by simply shifting the energy of mirror 8.741 MeV state in $^{22}$Ne by $\sim$130 keV. As shown in Fig.~\ref{fig2}(c), we assign
$J^{\pi}$=2$^{+}$ to this state. The observed 8.727 MeV state is regarded as the 8.743 MeV state of Matic {\it et al.}, which was
simply assumed to be the mirror of the 8.976 MeV, 4$^+$ state in $^{22}$Ne. The present $R$-matrix fit strongly prefers a 2$^{+}$
rather than a 4$^{+}$ as shown in Fig.~\ref{fig2}(d). It is worth mentioning that our data at the scattering angle of
$\theta_{c.m.}$$\approx$152${^\circ}$ also support the $J^{\pi}$ assignments discussed above~\cite{bib:zhang}.

\begin{figure}
\includegraphics[width=8.6cm]{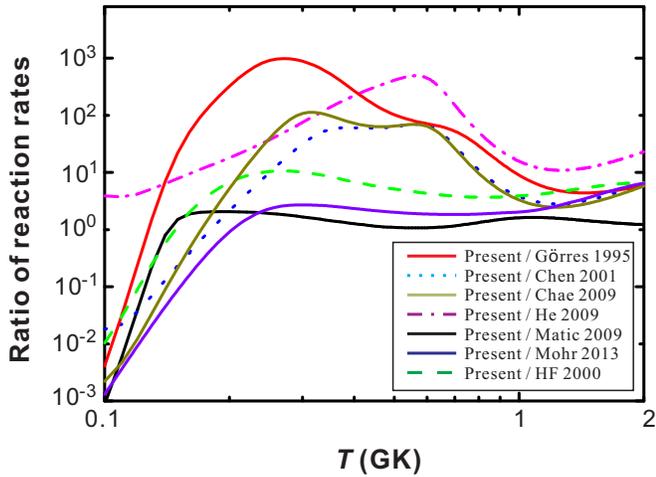}
\vspace{-6mm}
\caption{\label{fig3} (Color online) Ratios between the present rate and those previous ones (G\"{o}rres 1995~\cite{bib:gor95}, 
Chen 2011~\cite{bib:che01}, Chae 2009~\cite{bib:cha09}, He 2009~\cite{bib:he09}, Matic 2009~\cite{bib:mat09}, Mohr 2013~\cite{bib:mat13}, 
and statistical-model calculation HF 2000~\cite{bib:rau00}).}
\end{figure}

We have calculated the $^{18}$Ne($\alpha$,$p$)$^{21}$Na rate using a narrow resonance formalism~\cite{bib:che01,bib:mat09}. The
resonance parameters for the reaction rate calculations are summarized in Table~\ref{table1}. The proton partial widths ($\Gamma_p$)
deduced from our data will be given elsewhere~\cite{bib:zhang} as here we have calculated the $^{18}$Ne($\alpha$,$p$)$^{21}$Na rate
with $\omega\gamma$=$\frac{\omega\Gamma_\alpha \Gamma_p}{\Gamma_\mathrm{tot}}$$\simeq$$\omega\Gamma_\alpha$. The partial $\alpha$
widths are calculated~\cite{bib:gor95,bib:che01} by $\Gamma_{\alpha}$=$\frac{3\hbar^2}{\mu R_n^2}P_{\ell}(E)$$C^2S_{\alpha}$, where
$P_\ell$ is the Coulomb penetrability factor, $S_\alpha$ is the $\alpha$-spectroscopic factor, and $C$ is the isospin Clebsch-Gordon
coefficient. In this calculation, all resonance energies $E_R$ and most of the strengths $\omega\gamma$ are adopted from the work of
Matic {\it et al.}. For those states with new $J^\pi$ values determined in this work, the strengths are recalculated as listed in
Table~\ref{table1}. In this work, we adopt the 9.542 MeV state with a $J^\pi$=1$^-$ assignment as determined by an experimental
angular distribution measurement~\cite{bib:cha09}. The corresponding resonance strength is then about ten times larger than the value
in Ref.~\cite{bib:mat09}, where $J^\pi$=2$^+$ had been assumed. As for the 10.085 MeV state, we directly adopt the experimental
strength value~\cite{bib:gro02} rather than the calculated value from Ref.~\cite{bib:mat09}.
We note that experimental alpha spectroscopic factors ($S_\alpha$) for the states of concern in $^{22}$Mg are not available. As such, 
values (no uncertainties available) from mirror states in $^{22}$Ne have been adopted. Therefore, it is still difficult to 
quantitatively determine reliable uncertainties for the calculated rates.

The ratios between the present rate and the previous ones are shown in Fig.~\ref{fig3}. Comparing to the most recent rate of Matic
{\it et al.}, the present rate is much smaller below 0.13 GK. This is due to a unnatural-parity 1$^+$ newly assigned to the 8.385 MeV
state which does not contribute to the rate anymore. As well, the present rate is about 2.1 times larger around 0.2 GK, because of
our new $2^{+}$ assignments for the 8.657 and 8.743 MeV states. Finally, the present rate is about 1.6 times larger around 1.0 GK,
because of the experimental information we have adopted for the 9.542 and 10.085 MeV states as discussed above. Comparing to the
other available rates, our new rate is about a factor of 2--1000 times larger within the temperature region of interest for XRBs.
The curve labeled HF is a Hauser-Feshbach statistical-model calculation taken from Ref.~\cite{bib:rau00}. It shows that our new rate
is about 4--10 times larger than the theoretical prediction beyond 0.2 GK.

\begin{table}
\caption{\label{table1} The resonance parameters utilized for the $^{18}$Ne($\alpha$,$p$)$^{21}$Na rate calculation. Additional
energies and $J^{\pi}$ values determined in the present work, for states with $E_x$$<$8.2 MeV, are given in Fig~\ref{fig1}.}
\begin{ruledtabular}
\begin{tabular}{ccccccc}
 $E_x^{\mathrm{Pres.}}$ & $E_x^\mathrm{[18]}$ & $E_R^\mathrm{[18]}$ & $J^\pi$ & $S_\alpha$ & $\Gamma_\alpha$ & $\omega\gamma$ \\
 (MeV) & (MeV) & (MeV) &   &  & (eV) & (MeV) \\
\hline
8.190 & 8.181 & 0.039 & 2$^{+}$\footnotemark[1] & 0.284 & 1.7$\times$10$^{-65}$  & 8.53$\times$10$^{-71}$\footnotemark[2] \\
8.353 & 8.385 &       & 1$^{+}$\footnotemark[1] &       &                        &                                        \\
8.527 & 8.519 & 0.377 & 3$^{-}$\footnotemark[1] & 0.004 & 7.0$\times$10$^{-15}$  & 4.87$\times$10$^{-20}$\footnotemark[2] \\
8.578 & 8.574 & 0.432 & 4$^{+}$\footnotemark[1] & 0.06  & 3.6$\times$10$^{-13}$  & 3.26$\times$10$^{-18}$\footnotemark[2] \\
8.677 & 8.657 & 0.515 & 2$^{+}$\footnotemark[1] & 0.32  & 2.1$\times$10$^{-8}$   & 1.03$\times$10$^{-13}$\footnotemark[3] \\
8.727 & 8.743 & 0.601 & 2$^{+}$\footnotemark[1] & 0.11  & 2.7$\times$10$^{-7}$   & 1.34$\times$10$^{-12}$\footnotemark[3] \\
8.823 & 8.783 & 0.642 & 1$^{-}$\footnotemark[1] & 0.11  & 4.0$\times$10$^{-6}$   & 1.21$\times$10$^{-11}$\footnotemark[2] \\
8.932 & 8.932 & 0.790 & 2$^{+}$\footnotemark[1] & 0.11  & 8.3$\times$10$^{-5}$   & 4.13$\times$10$^{-10}$\footnotemark[2] \\
9.100 & 9.080 & 0.938 & 1$^{-}$\footnotemark[1] & 0.11  & 7.7$\times$10$^{-3}$   & 2.31$\times$10$^{-8}$ \footnotemark[2] \\
9.158 & 9.157 & 1.015 & 4$^{+}$\footnotemark[1] & 0.078 & 9.7$\times$10$^{-5}$   & 8.70$\times$10$^{-10}$\footnotemark[2] \\
                      & 9.318  & 1.176 & 2$^{+}$\footnotemark[4] & 0.11  & 9.9$\times$10$^{-2}$   & 4.97$\times$10$^{-7}$\footnotemark[2]  \\
                      & 9.482  & 1.342 & 3$^{-}$\footnotemark[4] & 0.015 & 1.8$\times$10$^{-2}$   & 1.25$\times$10$^{-7}$\footnotemark[2]  \\
                      & 9.542  & 1.401 & 1$^{-}$\footnotemark[5] & 0.11 & 5.24    & 1.57$\times$10$^{-5}$\footnotemark[3]  \\
                      & 9.709  & 1.565 & 0$^{+}$\footnotemark[4] & 0.15  & 5.2$\times$10$^{1}$    & 5.18$\times$10$^{-5}$\footnotemark[2]  \\
                      & 9.752  & 1.610 & 2$^{+}$\footnotemark[4] & 0.019 & 1.6                    & 8.22$\times$10$^{-6}$\footnotemark[2]  \\
                      & 9.860  & 1.718 & 0$^{+}$\footnotemark[4] & 0.019 & 2.1$\times$10$^{1}$    & 2.07$\times$10$^{-5}$\footnotemark[2]  \\
                      & 10.085 & 1.944 & 2$^{+}$\footnotemark[4] & --\footnotemark[6] & --\footnotemark[6] & 1.40$\times$10$^{-3}$\footnotemark[6]  \\
                      & 10.272 & 2.130 & 2$^{+}$\footnotemark[4] & --\footnotemark[6] & --\footnotemark[6] & 1.03$\times$10$^{-2}$\footnotemark[6]  \\
                      & 10.429 & 2.287 & 1$^{-}$\footnotemark[5] & --\footnotemark[6] & --\footnotemark[6] & 7.30$\times$10$^{-3}$\footnotemark[6]  \\
\end{tabular}
\end{ruledtabular}
\footnotetext[1]{Experimentally determined spin-parities in this work.}
\footnotetext[2]{All $S_\alpha$, $\Gamma_\alpha$ and $\omega\gamma$ values exactly adopted from Ref.~\cite{bib:mat09}.}
\footnotetext[3]{Recalculated $\Gamma_\alpha$ and $\omega\gamma$ values in this work.}
\footnotetext[4]{Spin-parities assumed in Ref.~\cite{bib:mat09} as adopted in the present work.}
\footnotetext[5]{Spin-parities determined in Ref.~\cite{bib:cha09}.}
\footnotetext[6]{Resonance strengths measured in Ref.~\cite{bib:gro02}.}
\end{table}

\begin{figure}
\includegraphics[width=8.6cm]{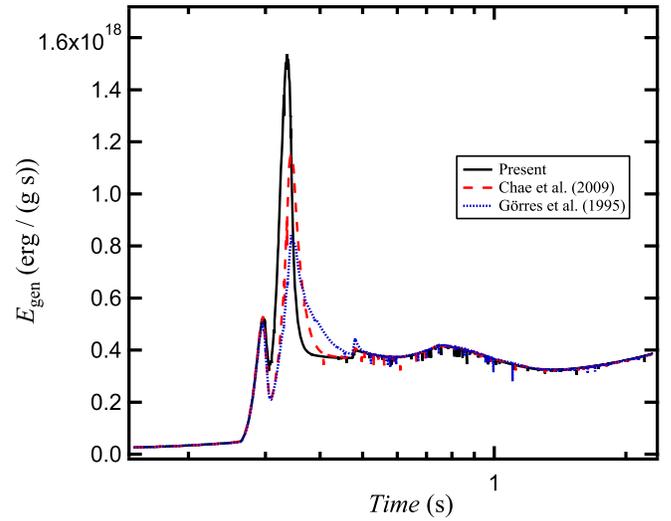}
\vspace{-6mm}
\caption{\label{fig4} (Color online) Nuclear energy generation rates during one-zone XRB calculations using the \texttt{K04}
thermodynamic history~\cite{bib:par08}. Results using the present rate (black solid line), the Chae {\it et al.} 2009~\cite{bib:cha09} 
rate (red dashed line) and the G\"{o}rres {\it et al.} 1995~\cite{bib:gor95} rate (blue dotted line) are indicated.}
\end{figure}

The impact of our new $^{18}$Ne($\alpha$,$p$)$^{21}$Na rate was examined within the framework of one-zone XRB postprocessing
calculations. Different XRB thermodynamic histories were employed, including the \texttt{K04} ($T_\mathrm{peak}$=1.4 GK) and
\texttt{S01} ($T_\mathrm{peak}$=1.9 GK) models from Refs.~\cite{bib:par08,bib:par09}. For each of these histories, separate
postprocessing calculations were performed using the present $^{18}$Ne($\alpha$,$p$)$^{21}$Na rate and previous
rates~\cite{bib:gor95,bib:che01,bib:mat09,bib:he09}; rates of all other reactions in the network~\cite{bib:par08} were left unchanged.

The rate of the $^{18}$Ne($\alpha$,$p$)$^{21}$Na reaction clearly affects predictions from our models. For example, as shown in
Fig.~\ref{fig4}, a striking difference in the nuclear energy generation rate at early times (between 0.3 and 0.4 s, or equivalently,
between 0.6 GK and 0.9 GK during the burst) is seen when comparing XRB calculations using the present, Chae {\it et al.} and
G\"{o}rres {\it et al.} rates with the \texttt{K04} model. Not only does the peak energy generation rate increase by a factor of
1.4--1.8 with the present rate, but the profiles of the curves around the maxima are also rather different. We also note a
change in the $^{18}$Ne($\alpha$,$p$)$^{21}$Na reaction flux at these early times. For example, at 0.35 s, this reaction flux
increases by a factor of 2--3 with our new rate. This contributes to the depletion of $^{15}$O and $^{18}$Ne at early
times by a factor of 3--4 relative to abundances calculated using the Chae {\it et al.} or G\"{o}rres {\it et al.}
rates.

We note that for both the \texttt{K04} and \texttt{S01} models, rates from Refs.~\cite{bib:gor95,bib:che01,bib:he09} give
lower peak nuclear energy generation rates than that from Chae {\it et al.}, by about 10--30~\%. Furthermore, the rate of
Matic {\it et al.} gives rather similar results to those using the present rate. In particular, the calculated nuclear energy
generation rates agree overall to about 5\%. This is of interest: despite the different $J^\pi$ values adopted in the present and
Matic {\it et al.} $^{18}$Ne($\alpha$,$p$)$^{21}$Na rate calculations (and the consequent differences in deduced thermonuclear
rates - see Fig.~\ref{fig3}), our models give very similar XRB nuclear energy generation rates. This suggests that $J^\pi$ values
for relevant states in $^{22}$Mg are, for the moment, sufficiently well known for our models. Future measurements should primarily
focus on measuring other quantities of interest (such as spectroscopic factors, partial widths or the precise direct cross-section
data), which can further constrain this rate.

In addition, with the present rate, the $^{18}$Ne($\alpha$,$p$)$^{21}$Na reaction dominates over the $\beta$-decay of $^{18}$Ne at
an onset temperature of $T$$\approx$0.47 GK (assuming a typical XRB density of 10$^6$ g/cm$^3$). This critical temperature is
noticeably lower than the breakout temperature of $T$$\approx$0.60 GK with the rates from Refs.~\cite{bib:gor95,bib:che01,bib:cha09},
and hence it implies that this reaction initiates the breakout earlier than previously thought.

Finally, we note that results from the XRB model adopted in Matic {\it et al.} are ostensibly in disagreement with our results.
When comparing calculated luminosities using their rate to that using the G\"{o}rres {\it et al.} rate, they found that a larger
$^{18}$Ne($\alpha$,$p$)$^{21}$Na rate may lead to a slightly lower peak luminosity. We find the opposite trend in our calculated
nuclear energy generation rates. Given this issue and the possible dramatic impact of the $^{18}$Ne($\alpha$,$p$)$^{21}$Na rate
in XRB models, it is clear that further tests using full hydrodynamic XRB models are needed to examine these effects in detail.

\begin{center}
\textbf{Acknowledgments}
\end{center}
We would like to thank the RIKEN and CNS staff for their friendly operation of the AVF cyclotron.
This work was financially supported by the National Natural Science Foundation of China (Nos. 11135005, 11021504), the Major State
Basic Research Development Program of China (2013CB834406), as well as supported by the JSPS KAKENHI (No. 21340053).
AP was supported by the Spanish MICINN (Nos. AYA2010-15685, EUI2009-04167), by the E.U. FEDER funds as well as by the ESF
EUROCORES Program EuroGENESIS.

\end{document}